\documentclass[pdflatex,sn-basic]{sn-jnl}

\usepackage{graphicx}%
\usepackage{amsmath,amssymb,amsfonts}%
\usepackage{amsthm}%
\usepackage{mathrsfs}%
\usepackage{xcolor}%

\let\orcidlogo\relax
\usepackage{orcidlink}

\begin{document}

\title[Accreting White Dwarfs: An Unreview]{Accreting White Dwarfs: An Unreview}


\author[1,2]{\fnm{Simone} \sur{Scaringi}\,\orcidlink{0000-0001-5387-7189}}\email{simone.scaringi@durham.ac.uk}

\author[3]{\fnm{Christian} \sur{Knigge}\,\orcidlink{0000-0002-1116-2553}}\email{c.knigge@soton.ac.uk}

\author[2]{\fnm{Domitilla} \sur{de Martino}\,\orcidlink{0000-0002-5069-4202}}\email{domitilla.demartino@inaf.it}

\affil[1]{\orgdiv{Centre for Extragalactic Astronomy, Department of Physics}, \orgname{Durham University}, \orgaddress{\street{South Road}, \city{Durham}, \postcode{DH1 3LE}, \country{United Kingdom}}}

\affil[2]{\orgdiv{INAF-Osservatorio Astronomico di Capodimonte}, \orgname{INAF}, \orgaddress{\street{Salita Moiariello 16}, \city{Naples}, \postcode{80131}, \country{Italy}}}

\affil[3]{\orgdiv{School of Physics and Astronomy}, \orgname{University of Southampton}, \orgaddress{\street{Highfield}, \city{Southampton}, \postcode{SO17 1BJ}, \country{United Kingdom}}}

\abstract{
Accreting white dwarfs (AWDs) are among the best natural laboratories for understanding disk accretion. Their proximity, brightness, and purely classical nature make them ideal systems in which to probe the fundamental physics that governs the transport of angular momentum, the generation of outflows, and the coupling between disks, magnetospheres, and accretors. Yet despite decades of study, many critical questions remain unresolved. In this ``unreview'', we therefore focus not on what is known, but on what is unknown. What drives viscosity and sustains accretion in largely neutral disks? How are powerful winds launched, and how do they feed back on the disk and binary evolution? Why do so many systems show persistent retrograde precession, and what drives bursts in magnetic AWDs? By identifying these open problems -- and suggesting ways to resolve them -- we aim to motivate new observational, numerical, and theoretical efforts that will advance our understanding of accretion physics across all mass scales, from white dwarfs to black holes.
}

\keywords{accretion, ejection, white dwarfs}

\maketitle

\section{Introduction}\label{sec1}

Accreting white dwarfs (AWDs) are arguably the best laboratories for studying accretion disks. First, they are the most common end-product of stellar evolution making them relatively bright and abundant in our Galaxy, with the closest system lying just 45 pc from the Sun \citep[e.g.][]{pala20}. Second, all of their key physical time scales -- dynamical, thermal, viscous, orbital, precession, eruption -- lie in the observationally accessible domain (seconds to years). Third, they are (or should be) relatively simple and purely classical systems with negligible general relativistic effects and a hard surface: if basic Shakura-Sunyaev disks exist anywhere, it should be in AWDs. Fourth (and in spite of the previous point), they exhibit almost the entire range of phenomenology seen in all kinds of accreting systems on all scales. This makes them an excellent sandbox for turning phenomenology into physics. In the language of computer programming: AWDs are nature's minimal reproducible example for debugging our understanding of accretion disks. 

In line with this, much of what we know about accretion disks has been learned from AWDs. For example, the interpretation of double-peaked emission lines as disk signatures was originally developed for and tested in AWDs \citep{smak81,HM86}. The same is true for both of the indirect disk imaging techniques in widespread use -- eclipse mapping \citep{horne85} and Doppler tomography \citep{MH88} -- which have led to breakthroughs such as the detection of spiral shocks \citep{steeghs01}. Similarly, the disk instability model \citep[DIM;][]{smak71,osaki74} that accounts for the transient nature of many neutron-star and black-hole X-ray binaries \citep[for a review see][]{lasota01} was initially designed to explain the behaviour of {\em dwarf novae} (DNe), a sub-class of AWDs that exhibit recurrent outbursts on time-scales of weeks to years \citep{smak71,osaki74,hoshi79,meyer81,smak82,smak84,cannizzo82,faulkner83,mineshige83}. Lastly, the inference of advection dominated accretion flows (ADAFs) at low mass transfer rates was first introduced in the context of the boundary layer in AWDs \citep{narayan93}, later continued for accretion disks in general \citep{narayan94}, eventually accounting for the comptonised component in many accreting neutron stars and black holes.

All of the above may seem like the obvious introduction to a review of all the things we (think we) know about accretion disks in AWDs. This is not that review. Instead, we have chosen to focus on what we {\em don't} know. As it turns out, there is rather lot of that. So what we will {\em actually} do is highlight what we think are some of the most fundamental open questions related to accretion disks in AWDs. Some of these are long-standing, and many of them have obvious and critical implications for disks in other types of accreting systems. For at least some of the questions, we will also sketch promising directions that might lead to an answer. And in case it is not obvious: our goal in compiling this list of problems is to motivate people smarter than us to solve them!

Throughout this ``unreview'', we will assume that the reader is familiar with the basic concepts of accretion disk theory. However, we will not assume familiarity with the extensive taxonomy and jargon associated with the various classes of AWDs. Thus we will take care to introduce any specialized terminology we rely on and avoid esoteric nomenclature -- like prototype-based class names (how about those ``IW And stars''?) -- wherever possible. That said, since the domain-specific literature will often use such nomenclature, we will generally mention it in passing. 

\begin{figure}[ht]
\centering
\includegraphics[width=0.9\textwidth]{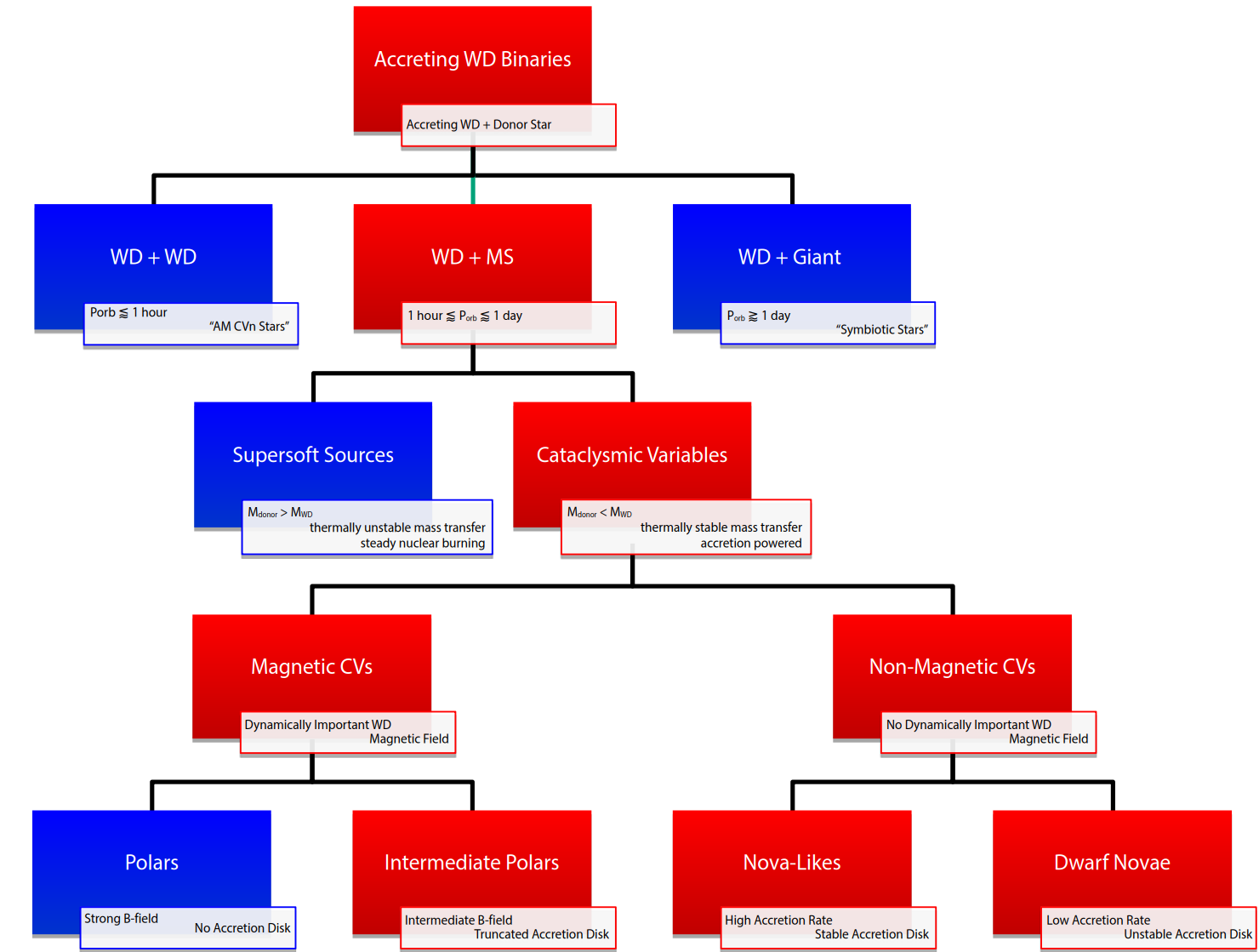}
\caption{Schematic taxonomy of accreting white dwarfs (AWDs). The diagram outlines the main subclasses: dwarf novae, nova-like variables, and intermediate polars. Systems discussed in this review are shown in red. These represent the ``vanilla'' disk-accreting white dwarfs with main-sequence donors and no steady nuclear burning, serving as benchmark laboratories for accretion physics.}
\label{fig:taxonomy}
\end{figure}

In this spirit, let us start with an overview of the diverse ``class'' of AWDs and outline what types of systems will -- and will not -- be covered here. The basic taxonomy of AWDs -- along with some of the relevant jargon -- is outlined in Figure~\ref{fig:taxonomy}. The systems shown in red in the bottom row are those we will focus on. These are generally referred to as Cataclysmic Variables (CVs) and might be called the most ``vanilla-flavoured'' of disk-accreting AWDs: those with roughly main-sequence donor stars, moderate accretion rates and no steady nuclear burning on the WD surface. We are restricting ourselves to these systems for two reasons. First, it makes the scope of this review at least somewhat manageable. Second, as we shall see, there are more than enough unsolved fundamental problems even among these ``well-understood'' systems.

As shown in Figure~\ref{fig:taxonomy}, our focus will be on three main sub-classes of AWDs, all of which have (roughly) main-sequence companions and orbital periods on the order of hours. The first and most famous are the above mentioned DNe. Observationally, these systems are characterized by the optical outbursts they exhibit, during which they brighten by factors of $\simeq 10 - 1000$. Depending on the system, these outbursts can have recurrence time-scales ranging from days to decades and last from $\simeq 1$~day to $\simeq 1$~month. Theoretically, this behaviour is interpreted as the consequence of the  thermal-viscous disk instability described by the DIM. Essentially, the mass-transfer rate from the donor star in DNe, $\dot{M}_{\rm 2}$, lies in a range where no stable thermal equilibrium is available for at least some parts of the disk. The physical reason for this is that the thermal equilibrium state for these regions would make them partially ionized -- but at the onset of Hydrogen ionization the opacity in this regime changes too fast and in the wrong direction. For example, any slight upwards fluctuation in the local mass-transfer rate, $\dot{M}_{\rm tr}$, and temperature makes these regions more ionized. This increases their opacity and drives the temperature and local mass-transfer rate up even more. Under the right conditions, this locally triggered thermal instability can then spread viscously across the disk. DN disks thus cycle between a cool, faint, mostly neutral quiescent state with $\dot{M}_{\rm tr} < \dot{M}_2$ and a hot, bright, mostly ionized outburst state with $\dot{M}_{\rm tr} > \dot{M}_2$ \citep{lasota01}.

The second class of AWDs we will focus on are the ``nova-like'' variable (NLs). In these systems, $\dot{M}_2$ is thought to be high enough to keep the disk globally and permanently in the ionized state, so it is not susceptible to the DN instability. NLs thus do not exhibit the recurrent outbursts characteristic of DNe. They are (or should be) the closest approximation we have to a simple, steady-state, optically thick, geometrically thin \cite{SS73} disk.

In both DNe and NLs, the magnetic field of the WD is usually assumed to be dynamically insignificant -- it does not affect the accretion flow (but as we shall see, this may not be a safe assumption in many cases). In the jargon of the field, DNe and NLs are ``non-magnetic CVs'' (although suspected magnetic WD systems sometimes tend to be referred as NLs before the proper magnetic classification). This contrasts with the third class of disk fed AWDs we will cover, the ``intermediate polars'' (IPs). In these systems, the magnetic field of the WD is strong enough to truncate the accretion disk at radii $R_{\rm trunc} > R_{\rm WD}$. Material is then expected to be channelled into ``accretion curtains'' and flow along field lines onto the magnetic poles of the WD \citep{rosen88,hellier95}.

\section{Which angular momentum transport (or loss) mechanisms drive disk accretion in AWDs?}
\label{sec:visc}

Let us start with the elephant in the room: the mechanism that actually drives accretion through the disk. The key physics here is simple: the angular momentum of a test particle in a circular Keplerian orbit around a gravitating mass increases with radius as radius $R^{1/2}$. Thus in order for material to move from large to small radii in the disk -- i.e. in order for accretion to occur at all -- it must lose angular momentum. Fundamentally, there are two ways in which this might happen: (i) angular momentum might be {\em extracted from the disk} via some sort of outflow; (ii) angular momentum might be {\em redistributed within the disk} by some form of viscosity.

Since normal collisional (``molecular'') viscosity is far too inefficient to drive the observed accretion rates in astrophysical systems, \cite{SS73} (hereafter SS73) parametrised the disk viscosity as $\nu = \alpha \, H \, c_{\rm s}$, where $\alpha$ is a dimensionless fudge factor characterizing the strength of viscosity, $H$ is the vertical scale height of the disk, and $c_{\rm s}$ is the local sound speed. In adopting this prescription, SS73 anticipated that disk accretion would ultimately be driven by some sort of {\em turbulent} viscosity. For this type of viscosity, $\alpha \lesssim 1$, since  (isotropic) turbulent eddies cannot be much larger than the disk thickness, and their turnover speed cannot exceed the sound speed (since otherwise shocks would be formed). 

Over the last few decades, the magnetorotational instability \citep[MRI;][]{balbus_hawley91} has become the dominant paradigm for angular momentum transport in most types of disk-accreting astrophysical systems, including AWDs. The basic physical picture here is that even weak magnetic fields can provide ``spring-like'' connections between parcels of gas in the disk. Such parcels therefore resist being pulled apart by differential rotation, providing the effective viscosity that can then drive accretion. Note that this requires the parcels to be sufficiently ionized, as otherwise they will be largely indifferent to the presence of magnetic fields.

However, attempts to model AWD disks within this framework have encountered both quantitative and qualitative challenges. One important issue is related to the magnitude of $\alpha$. The viscosity parameter appropriate to DNe in outburst (i.e. to hot, ionized disks), $\alpha_{\rm o}$, can be estimated empirically from the observed rates of decline. Such estimates typically give $\alpha_{\rm o} \simeq 0.1 - 0.3$ \citep{smak99,kotko12}. There are no comparably robust ways to estimate the viscosity parameter in quiescence (i.e. in cool, mostly neutral disks), $\alpha_{\rm q}$, although \cite{king13} have argued that $\alpha_{\rm q} \lesssim 10^{-4}$ is required to explain the persistence of tilted disks across the entire quiescence-outburst cycle in at least some DNe \citep{harvey95}. These estimates are broadly in line with the DIM, which requires that the viscosity parameter should be significantly higher in outburst ($\alpha_o \gtrsim 0.1$) than in quiescence ($\alpha_{\rm q} \lesssim 0.01$). If this requirement is not met, the locally triggered thermal instability will not spread fast enough to trigger a {\em global} disk eruption. 

Magnetohydrodynamic (MHD) simulations that resolve the MRI have long struggled to achieve the high angular momentum transport efficiencies inferred for DNe in outburst. In such simulations, the viscosity rarely exceeds $\alpha_{\rm h} \simeq 0.01$, an order of magnitude or so less than required \citep[e.g.][]{king07,kotko12}. How serious is this problem? It is difficult to be sure, because the efficiency of angular momentum transport in these simulations is sensitive to wide range of physical and numerical factors. These include the dimensionality of the simulations (2-D vs 3-D), the pseudo-periodic boundary conditions that are usually adopted, the vertical resolution (the number of zones used to resolve the disk scale height), whether the accretion stream is included (which can excite spiral shocks that can also drive angular momentum transport; see below), the strength and geometry of the initial seed magnetic field (stronger and more vertical fields tend to produce higher $\alpha$) and the treatment of radiation and thermodynamics (e.g. what is assumed about cooling, the equation of state, the scale height and the Mach number). As a result, even though there {\em are} MHD simulations that have achieved values of $\alpha_{\rm o}$ closer to what is observed \citep[e.g.][]{sorathia12,hirose14,ju17}, it is unclear whether the MRI can or cannot account for high-state AWDs (i.e. DNe in outburst and NLs). 

The situation is even worse for low-state AWDs, i.e. DNe in quiescence. The angular momentum efficiency required to explain these systems is much lower, but accretion does take place in these disks. We know this because quiescent DNe do produce persistent X-ray emission -- more than they should (although in some cases at lower fluxes than expected from boundary layer models \citep{belloni91,wheatley96,Pandel03}. This means AWDs are fed by disk material even in quiescence. But should the MRI be able to operate at all in these disks, given that they are expected to be largely neutral? The answer seems to be ``probably not'' \citep[e.g.][]{gammie98,scepi18a}, and we note this challenge is also one to be faced by disks in other types of accreting systems.

Given these challenges to MRI-driven accretion in both low-state and high-state systems, let us briefly consider two of the most promising alternative (or additional) mechanisms: wind-driven accretion and accretion driven by spiral shocks. 

The possibility that angular momentum might be {\em extracted} from the disk by an outflow (rather than being {\em redistributed} by viscosity) was already alluded to above. This idea has been around for some time \citep{cannizzo88,livio94}, but primarily in the context of the secular evolution of AWDs as binary systems (which depends critically on the systemic angular momentum loss rate). Here, we just want to highlight a few key points regarding such wind-driven disks that we think are not always appreciated. First, only {\em magnetic} disk winds can drive accretion. Radiatively- or thermally-driven outflows do, of course, also extract mass and angular momentum from the disk. However, since they are expected to conserve specific angular momentum, they effectively only extract the angular momentum of the ejected material itself. This does not exert a braking torque on the remaining disk material, i.e. it does not allow this material to spiral inwards. By contrast, ionized material launched by a magnetic disk wind from cylindrical radius $R_{\rm l}$ must initially move along the (poloidal) field lines, It thus effectively corotates with $R_{\rm l}$ out to the Alfv\'{e}n radius, $R_{\rm A} > R_{\rm l}$, where gas pressure begins to dominate over magnetic pressure and co-rotation breaks down. Magnetic winds {\em do} exert a braking torque on the remaining disk material, since the wind material carries away {\em more} specific angular momentum, $j_{\rm w} = R_{\rm A}^2 \Omega_K(R_{\rm l})$ (where $\Omega_K(R)$ is the Keplerian frequency at radius $R$), than it had at the launch radius on the disk, $j_{\rm l} = R_{\rm l}^2 \Omega_K(R_{\rm l})$. The magnitude of this torque depends on the effective ``lever arm'' of the wind, $l = R_{\rm A}/R_{\rm l}$. 

Second, a purely wind-driven disk would be cold and completely dark \cite[e.g.][]{spruit95,knigge99,scepi19}. This is easy to understand: in a viscously-driven disk, the gravitational potential energy of the infalling material is first converted to internal energy in the disk and then radiated away. But this conversion takes place via viscous dissipation -- without viscous angular momentum transport, there is also no viscous heating and no radiation. In any wind-dominated accretion scenario, the radiative luminosity of the disk must be far less than the true accretion luminosity, $L_{\rm rad, disk} << L_{\rm acc}$. We suspect this rules out wind-driven accretion at least in high-state AWDs: if their accretion rates were much higher than implied by the radiative luminosities, their WDs would likely be in the steady nuclear burning regime (i.e. they would be supersoft sources; c.f. Figure~\ref{fig:taxonomy}). 

Third, wind-driven accretion may be more promising for quiescent DNe \citep{scepi18,scepi19}. At first sight, this is counter-intuitive: the MRI is quenched here because the disk becomes largely neutral, but neutral material can also not be forced to corotate with the field lines of a magnetic disk wind. The subtle resolution here is that, while the bulk of the disk may be neutral, a thin upper layer is likely to be ionized (by irradiation from the WD or boundary layer/corona; see Section~\ref{sec:hotFlow}). In this layer, both the MRI and magnetic disk winds may be able to operate. However, if only material in this layer can  accrete, the inward radial velocity has to be supersonic to match the required accretion rates. This is impossible for turbulent, MRI-driven viscosity, but not for wind-driven accretion.

Fourth, all of the observational evidence for disk winds in AWDs comes from high-state systems \citep[NLs and DNe in outburst; c.f. Section~\ref{sec:visc};][]{matthews15}. In the present context, this is a little frustrating: we only see evidence for winds in systems where wind-driving cannot work. However, absence of evidence is not evidence of absence. The kinds of wind signatures we are used to seeing in high-state systems require relatively high mass-loss rates and densities. Magnetic disk winds might exist in quiescent DNe, but simply not produce these signatures. Observational searches for disk winds in quiescent DNe -- perhaps via weakly eclipsed wind-formed emission lines in high-inclination systems \citep[c.f.][]{horne94} -- are strongly encouraged. 

The other promising alternative mechanism for driving accretion in the disks of AWDs are spiral shocks \citep[e.g.][]{boffin01,ju16,ju17}. Such shocks are expected, since the gravitational potential is non-axisymmetric (due to the donor star) and since matter is injected into the disk at a specific location (the so-called ``hot spot'' where the accretion stream meets the disk edge). There is a long history of theoretical \citep[e.g.][]{spruit} and numerical \citep[e.g.][]{sawada} work on these structures, and they have the distinct advantage of actually having been observed \citep[in at least some systems: see, e.g.][]{steeghs97,steeghs00,ruiz20a,ruiz20b,pala19}.

\begin{figure}[ht]
\centering
\includegraphics[width=0.9\textwidth]{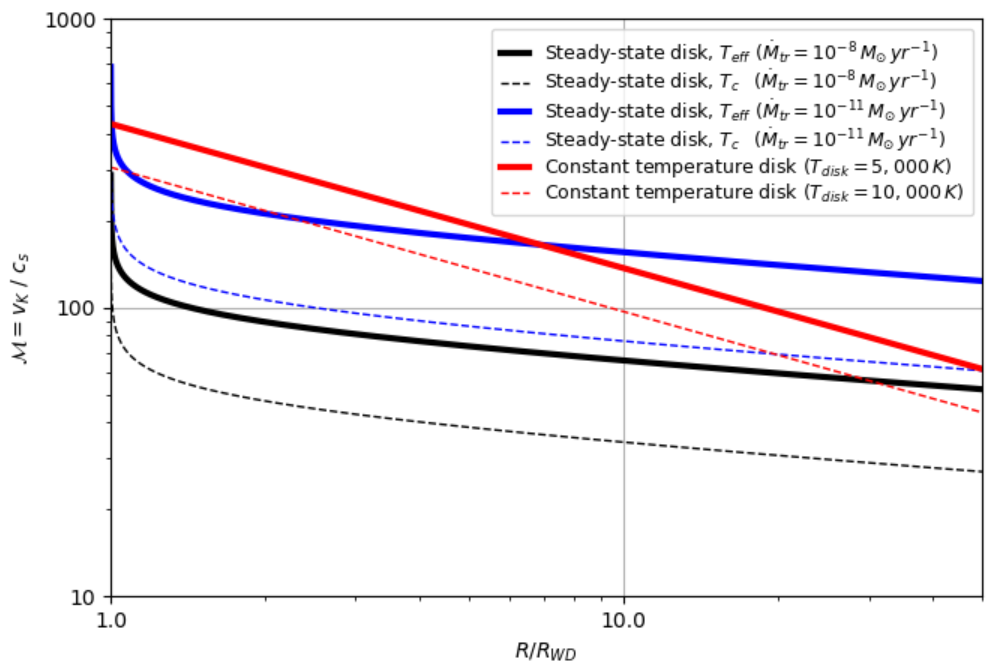}
\caption{The Mach number as a function of disk radius in several simple accretion disk models with parameters appropriate for AWDs. For the steady-state (Shakura-Sunyaev) disks, we show the Mach numbers corresponding to both the central and effective disk temperature. In the isothermal models, the temperature is assumed to be constant throughout the disk.}
\label{fig:mach}
\end{figure}

A key challenge for spiral-shock-driven accretion is that it is only viable at low Mach numbers ($\mathcal{M} = v_{\rm \phi}/c_{\rm s} \lesssim 10$; \citep[e.g.][]{blondin00,ju16,ju17}. For high Mach numbers, the shocks become weaker and tightly wound -- more so than observed and too much so to transport angular momentum efficiently. Figure~\ref{fig:mach} shows characteristic Mach numbers for Keplerian accretion disks in AWD systems. The sound speed increases with temperature as roughly $c_s \simeq 10~(T/10^4~{\mathrm{K}})^{(1/2)}$, so hot, high-$\dot{M}_{\rm tr}$ disks have lower Mach numbers. However, Mach numbers $\mathcal{M} \gtrsim 30$ are expected even near the mid-plane of the outer disks in high-state AWDs (NLs and DNe in outburst). This makes it very hard to sustain strong spiral shocks in {\em any} AWD disk, let alone quiescent DNe. That said, as was the case with the MRI, 3-D global, radiation-magneto-hydrodynamic simulations will be needed to assess the viability of spiral-shock-driven angular momentum transport in accretion disks.

Where does this leave us? It seems to us that the question of what drives accretion in AWDs -- both in outburst and in quiescence -- is far from settled. This should concern anybody interested in accretion physics -- if we don't even understand disk accretion in AWDs, the simplest and cleanest such systems, many applications of disk theory to more complex systems must also be suspect. We therefore strongly encourage numerical (magneto-)hydrodynamicists to renew their focus on AWDs. In one sense, AWDs are low-hanging fruit for realistic, global 3-D simulations: the radial dynamic range one has to cover is small (their disks typically span less than 2 orders of magnitude) and non-relativistic ideal MHD is an excellent approximation. Instead, the challenge their disks pose is that they really are {\em thin} disks. This makes them very hard to resolve in the vertical direction, and doing so requires a realistic treatment of energy transport via radiation. However, computational techniques and resources may finally have a reached a point where this is not an insurmountable challenge anymore. We therefore hope that ambitious and enterprising theorists will rise to it.

\section{How do accretion disk winds fit into our physical picture of AWDs?}
\label{AWD_winds}

\begin{figure}[ht]
\centering
\includegraphics[width=0.3\textwidth]{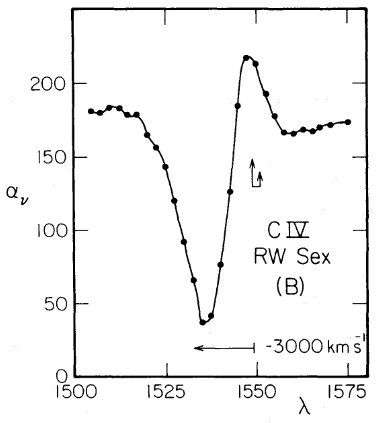}
\caption{The C~{\sc iv} $\lambda\lambda 1550$ line in the NL variable RW~Sex. This P~Cygni profile, with its strong, blue-shifted absorption, was one of the first clear indications that high-state AWDs must drive powerful disk winds \citep[Figure credit:][]{greenstein82}}
\label{fig:pcyg}
\end{figure}

We have known for over four decades that high-state AWDs -- i.e. NLs and DNe in outburst -- experience mass loss, probably in the form of accretion disk winds \citep{heap78,krautter81,klare82,greenstein82,cordova82}. As shown in Figure~\ref{fig:pcyg}, the original discoveries were based on the detection of blue-shifted absorption or P~Cygni profiles in the ultraviolet (UV) resonance lines, such as N~{\sc v}~$\lambda\lambda 1240$, Si~{\sc iv}~$\lambda\lambda 1400$ and C~{\sc iv}~$\lambda\lambda 1550$ \citep[also see, for example,][]{ladous}. This is a smoking gun for the presence of mass loss.

However, despite the long-standing recognition that the disks in (at least) high-state AWDs must drive outflows, our {\em understanding} of these winds remains poor. For example, we do not even know the basic driving mechanism with any certainty. It is, of course, important to ask if this matters. If the observational and physical impact of disk winds is limited to the production of P-Cygni profiles in a few UV lines, perhaps we don't need to worry about understanding them. Unfortunately, life is not so simple: disk winds may actually have a significant impact on everything from the accretion process itself, to the evolution of AWDs as binary stars, to the observational appearance of these systems across the entire spectral range. 

Before discussing how and why disk winds are relevant to these areas, let us very briefly review the three fundamental mechanisms that can drive mass-loss from AWDs (and, in fact, any astrophysical system): (i) thermal driving; (ii) radiative driving; (iii) magnetic driving. 

{\em Thermal driving} refers to the production of outflows via thermal expansion. Mass loss from a system is inevitable whenever the thermal speed of (some) particles in a medium exceeds the local escape speed. This mechanism is fundamentally responsible for the solar wind, and it is also likely to be responsible for driving the disk winds in at least some X-ray binary systems \citep[e.g.][]{done07,higginbottom17,higginbottom20,xrismcoll25}. Thermal driving has been considered in the context of AWDs \citep{czerny89a,czerny89b,deKool89}, but -- at least for high-state systems -- it is difficult to see how the top of the accretion disk atmosphere would be heated to the required temperatures ($T \gtrsim 10^6$~K). 

{\em Radiative driving} refers to the momentum transfer associated with scattering and/or reprocessing of photons. The simplest example of such an interaction is electron scattering, and the famous Eddington limit corresponds to the maximum luminosity a (spherical) self-gravitating object can have before the outward force associated with electron scattering will overcome the inward gravitational pull. Even high-state AWDs have luminosities far below the Eddington limit, however, so if their winds are radiatively driven, other types of interactions must dominate. Since the physical conditions in these systems are not too far from those found in hot stars -- in terms of characteristic temperatures and surface gravities -- the specific mechanism usually invoked in this context is radiation pressure mediated by spectral lines (``line driving''). Indeed, {\em some} of this {\em must} be happening: after all, C~{\sc iv} profiles like that in Figure~\ref{fig:pcyg} are produced by scattering events in which momentum is transferred from the radiation field to the outflow.

{\em Magnetic driving} refers to the action of either magnetic pressure gradients and/or (more commonly) centrifugal forces when ionized material is forced to co-rotate with field lines anchored in the accretion disk and extending to larger cylindrical radii. The classic model for magneto-centrifugal disk winds was developed by Blandford \& Payne (1982), who also introduced the useful analogy of ``beads sliding outwards on a rotating wire'' for this mechanism. As already noted in Section~\ref{sec:visc}, magnetic driving is the {\em only} mechanism that naturally extracts ``extra'' angular momentum from the accretion disk (in the sense of actually exerting a torque on the material remaining in the disk). The main requirement for magnetic driving of this kind is the existence of a sufficiently strong field with the ``right'' kind of geometry. In particular, the classic \cite{BP82} picture requires field lines to be inclined at least 30$^{\circ}$ from the disk normal.

In high-state AWDs, line-driving has long been the leading candidate for generating the observationally inferred outflow \citep{drew85,mauche87,vitello88,hoare93,mauche94,drew97,feldmeier99a,feldmeier99b}. This is largely based on the physical and observational similarities between the outflows from AWDs and those of hot stars (which are known to be line-driven). However, 
there are serious challenges to this interpretation. Observationally, one significant problem is that one might expect the mass-loss rate -- and hence the strength of wind features -- to scale with luminosity in a radiatively-driven wind. In the two systems where this has been tested directly via time-resolved spectroscopy no such correlation was found \citep{hartley02}. 

There are also theoretical problems. The most fundamental issue is that AWDs are only {\em marginally} capable of launching line-driven winds at all. For the sort of conditions found in the winds of hot stars and expected in those of AWDs, atomic physics essentially limits the enhancement spectral lines can provide over the radiation pressure force exerted purely by electron scattering. This enhancement factor is known as the ``force multiplier'', $\mathcal{M}$, and -- for early type stars with SEDs similar to NLs -- the maximum value it can take is around $\mathcal{M}_{\rm max} \simeq 2000$ \citep[e.g.][]{gayley95}. In high-state AWDs, $L_{\rm acc}/L_{\rm Edd} \simeq 10^{-3}$, i.e. $L_{\rm acc}/L_{\rm Edd} \simeq \mathcal{M}_{\rm max}^{-1}$. Thus line-driving can only work in AWDs under near optimal conditions.

Most radiation-hydrodynamic simulations carried out to date have relied on highly approximate, quasi-1D treatments of ionization and radiative transfer \citep[e.g.][]{pereyra97,proga98,proga99,pereyra00, dp1, dp2, dp3, dp4}. These treatments amount to effectively {\em assuming} that near optimal conditions exist in the wind-driving region \citep{sim10,higginbottom14}. Yet even with these optimistic assumptions, the mass-loss rates and observational signatures produced in these simulations were significantly weaker than observed \citep[e.g.][]{drew00}. Simulations in which radiative transfer and ionization are treated self-consistently only became possible recently -- and the first results were just as bad as one might have expected. The ionization state of the outflows in these state-of-the-art simulations is much higher than that in earlier simulations. This shift changes the set of bound-bound transitions that can drive the outflow, making line driving less effective. The upshot is that the outflows generated in these simulations are extremely weak \citep{higginbottom24,mosallanezhad25}.

Lest all of this sound too gloomy, there are good reasons to hope that these challenges can be resolved. First, we know observationally that the conditions in AWD disk winds {\em are} nearly optimal -- because the UV wind signatures we know and love (e.g. Figure~\ref{fig:pcyg}) {\em require} these conditions. Indeed, as noted above, these signatures themselves  contribute to line driving, so line driving {\em must} be happening to some degree. If simulations of line-driven winds fail to produce the observationally inferred conditions, the problem may lie in the simulations, not the mechanism itself. Second, there is at least one physically plausible get-out-of-jail-free card: clumping. The 
line-driven winds of hot stars are known to be highly structured, probably due to an inherent instability associated with the mechanism -- the so-called ``line-deshadowing instability'' \citep[see, for example, the review by ][]{Puls}. This instability sets in on scales that are too small to resolve in even the most ambitious state-of-the-art global simulations. However, we {\em can} use a simple sub-grid method -- ``micro-clumping'' \cite[e.g.][]{hamann, matthews_clumpy} -- to test the effect of clumping on the wind ionization state and the resulting line forces. Such simulations show that even moderate clumping -- corresponding to filling factors of a few percent -- are sufficient to lower the ionization state of the outflow to the point where line driving can launch powerful outflows (Mosallanezhad et al. 2025, in prep).

What about {\em low-state} AWDs, i.e. DNe in quiescence? Line driving {\em cannot} work in these systems ($L/L_{\rm Edd} << \mathcal{M}_{\rm max}^{-1}$), so if they do produce significant mass loss, thermal or magnetic driving must be responsible. As already discussed in Section~\ref{sec:visc}, magnetic driving may be particularly attractive in this context. This is because the MRI is likely to be quenched in quiescent disks, so magnetic wind torques might be needed to drive accretion through these disks at all. 

The possibility that disk winds might be {\em required} to drive accretion is one important reason to care about them. But there are at least two others. First, if a magnetic outflow were to extract significant amounts of angular momentum from the disk -- and thus perhaps drive accretion -- the angular momentum carried away by the wind would also be lost from the binary system. This could have a major impact on the  secular evolution of AWDs, which is entirely driven by angular momentum losses \citep[AML; see][for a comprehensive review]{knigge11}. Briefly, in the ``classic'' evolutionary scenario for AWDs, AML is dominated by ``magnetic braking'' associated with the donor star at long orbital periods ($P_{\rm orb} \gtrsim 3$~hrs) and by the emission of gravitational radiation at short-periods ($P_{\rm orb} \lesssim 2$~hrs). However, there is strong evidence that, at least for short-period systems, an additional AML mechanism must be at play. ``Magnetic braking'' associated with a disk wind might play such a role \citep{cannizzo88,livio94}. In the context of binary evolution, a disk wind is an example of a ``consequential'' AML mechanism, in the sense that it can only {\em amplify} an existing mass-transfer mechanism. After all, it requires a disk to already be present in the first place.

The second reason we should care about disk winds in AWDs is that they might have a dramatic -- and underappreciated -- effect on observations across a wide range of wavelengths. For example, P-Cygni profiles are also quite commonly seen in optical recombination lines \citep[e.g.][]{kafka04}, and the single-peaked emission lines seen in many NLs are also likely to be wind-formed \citep[e.g.][]{murray96,murray97}. In fact, it is plausible that most or all of the optical emission lines seen in high-state AWDs are formed in the dense base of a disk wind \citep[e.g.][]{knigge97a,matthews15,tampo22,tampo24,wallis25}. 

It is not just the spectral {\em lines} that might be affected, however. The base of the wind may actually reprocess a significant fraction of the luminosity emitted by the disk, in which case it could dramatically change the overall spectral energy distribution. This type of reprocessing might explain some long-standing observational problems. For example, absorption in the outflow (rather than just the interstellar medium [ISM]) may account for the ``missing boundary layer'' in AWDs \citep{ferland82,kallman85,hoare93,long94}. Moreover, the re-emission of this absorbed radiation naturally produces a recombination continuum that could go a long way towards explaining the flatter-than-expected UV/optical continua and weaker-than-expected absorptive Balmer jumps in high-state AWDs \citep[e.g.][]{wade84,long91,long94,knigge97a,knigge98a,matthews15}. Another observational signature of disk winds are the spectacular bow shocks they can produce when they interact with the ISM \citep[e.g.][]{krautter87,hollis92,miszalski16,bond18,hernandez19,castro21,bond24,bond25}. 

So, disk winds matter. We know they are present in at least high-state AWDs. They may act as angular momentum sinks that drive both accretion and binary evolution. And they can dramatically affect our view of the underlying system. If we want to understand accretion disks, we have to understand their disk winds as well.

How can we make progress in this area? A good start would be to identify the wind driving mechanism in high-state AWDs with more confidence. The clumpy line-driving simulations presented in \cite{mosallanezhad25} are a significant step in this direction. Pushing this type of modelling beyond the proof-of-concept stage should be a high priority. It is also important to construct a better observational picture of these outflows -- or at least a less model-dependent one. For example, bow shocks can be used to obtain robust estimates of the wind mass-loss rates \citep[e.g.][]{hollis92,castro21}. Similarly, simple parametrised disk wind models can be used to model observed spectra to infer key aspects of the outflow geometry and kinematics \citep[e.g.][]{knigge97a,knigge97b,noebauer10,wallis25}. By combining as many types of observational constraints across as many systems as possible, we can iterate towards a robust physical picture and identify trends with system parameters. For example, is the efficiency of the outflow -- $\dot{M}_{\rm wind} / \dot{M}_{\rm acc}$ -- a function of accretion rate? White dwarf mass? Disk radius? Donor star properties?

\section{Do accreting white dwarfs drive (radio) jets?}

The previous section was concerned with disk winds, but another type of outflow is also common among disk-accreting astrophysical systems: jets. The primary distinction between these two types of outflows is collimation: disk winds are relatively poorly collimated (think opening angles $\Delta \theta$ spanning tens of degrees), jets are highly collimated (think $\Delta \theta \lesssim 10^{\circ}$). Many disk-accreting astrophysical systems drive {\em both} kinds of outflow, including young stellar objects \citep[YSOs; e.g.][]{bally16}, X-ray binaries \citep[XRBs; e.g.][]{fender14} and active galactic nuclei \citep[AGN; e.g.][]{morganti17}. 

In most cases, there are additional distinctions. Disk winds are typically more highly mass-loaded, but jets are faster (in XRBs and AGN, the jets are relativistic). Their radiative signatures are also different: the impact of disk winds on the spectra of AWDs was already discussed in the previous section, and similar signatures are seen in other wind-driving systems. By contrast, the primary signature of jets often tends to be radio emission via synchrotron or free-free radiation.

Jet formation mechanisms invariably involve magnetic fields. More specifically, most jet launching scenarios rely on either the Blandford-Znajek \citep[BZ;][]{BZ} effect or the Blandford-Payne \citep[BP;][]{BP} magneto-centrifugal mechanism (c.f. Section~\ref{sec:visc}). The BZ effect requires the presence of a rotating black hole and is therefore not relevant to this Chapter on AWDs. The obvious question one might ask regarding the BP mechanism -- which is thought to be responsible for the non-relativistic jets in YSOs \citep[see][]{meier97}, for example -- is ``why would a magneto-centrifugally launched outflow be highly collimated?''. After all, as noted in Section~\ref{AWD_winds}, the mechanism requires the magnetic field to be inclined by {\em at least} 30$^{\circ}$ relative to the disk normal. The answer is hoop stresses: once a magnetically launched wind reaches the Alfv\'{e}n point, where gas pressure begins to exceed magnetic pressure, the field lines are forced to move with the gas. They are therefore wound up azimuthally, turning the initially poloidal field into a mostly toroidal (and very strong) field. It is the magnetic pressure associated with this toroidal field well above the Alfv\'{e}n surface that collimates the outflow. We can incorporate this into the ``bead-on-a-rotating-wire'' analogy by not  allowing the wire to be infinitely rigid. The Alfv\'{e}n point then corresponds to the location where the wire begins to be bent by the beads sliding along it. 

The previous paragraph illustrates that the relatively clear {\em observational} distinction between disk winds and jets we have outlined above does not necessarily translate into such a clear {\em physical} or {\em theoretical} distinction. A BP-type outflow may be poorly collimated and relatively slow initially (i.e. a disk wind), but might then accelerate and become highly collimated by hoop stresses far from the disk (i.e. it may turn into a jet). It is therefore not surprising that the {\em connection} between disks, jets and winds is one of the most important open questions in accretion physics quite generally.

But what does any of this have to do with AWDs? Quite a lot, as it turns out. In fact, the supposed {\em absence} of jets from AWDs was used by \cite{livio97,livio99} to argue that jets require the presence of a strong central power source -- something that might be missing in AWDs, but present in other kinds of disk-accreting systems. This originally prompted unsuccessful searches for {\em optical} jet signatures in AWDs \citep[e.g.][]{shahbaz97,margon98,obrien98,knigge98b,hillwig04}. However, almost two decades ago now, \cite{koerding11} detected radio emission associated with an erupting DN. This discovery wasn't a fluke: the observations had been carefully timed to catch the kind of radio flare that is commonly seen in XRBs during their outbursts. The radio flares in these systems are known to be associated with powerful transient jets, and the material ejected during these episodes can often be directly imaged and tracked for days or weeks after the event \citep[e.g.][]{fender04}. The discovery of a similar kind of flare in the famous DN system SS~Cyg was therefore immediately interpreted as strong evidence for the presence of radio jets in AWDs as well. 

Since then, the picture has become significantly murkier (perhaps we should say "even more interesting"). On the one hand, radio emission has now been detected in a significant sample of both NLs and DNe \citep[e.g.][]{coppejans15,coppejans16,kersten05}. This makes it quite tempting to assume that (radio) jets must be present in these systems. On the other hand, the devil is very much in the details. For example, unlike in XRBs, the radio emission in DNe does {\em not} appear to be fully quenched after the radio flare. In fact, the most detailed observation of a single (anomalous) outburst in SS~Cyg revealed dramatic (and completely unexpected) flaring activity throughout \citep{mooley17}. Yet the overall radio behaviour of the system across outbursts appears to be quite reproducible \citep{russell16}. There is also at least one DNe that does {\em not} seem to produce radio emission at the expected level at all \citep[VW~Hyi; see][]{coppejans00}. The situation is no clearer among the NLs, which exhibit disturbingly heterogenous variability properties, spectral indices and polarization levels. 

Since there is already a recent and comprehensive review of the case for (and against) jets in AWDs \citep{coppejans00}, we will end this section here with just three additional remarks. First, we hope our discussion has made it clear that settling this question is important -- not just for AWDs, but for our understanding of accretion physics more generally. Second, we are not aware of a single (magneto-)hydrodynamic simulation aimed at describing jet launching (e.g. via the BP mechanism) from an AWD. It is high time to rectify this. Third and finally, while there are still many gaps in the observational picture, the lack of a {\em spatially resolved image} of a jet in an AWD is clearly the most important -- it would be the smoking gun confirming that AWDs  {\em do} drive jets. Of course, there is no guarantee that such a detection is possible with present-day instrumentation, even if jets do exist in AWDs. However, the VLBA image of SS~Cyg presented by \cite{russell16} actually did exhibit marginal evidence for a resolved jet. This provides extremely strong motivation for additional,  deeper VLBI observations that might confirm this hint.

\section{What triggers and sustains the misalignment and retrograde precession of disks in AWDs?}\label{sec:tilt}

Accretion disks in CVs are most often treated as planar, geometrically thin structures that remain aligned with the binary orbital plane under the gravitational influence of the donor star. However, since the first detection of optical negative superhumps in a handful of high mass transfer rate NLs \citep[e.g.][]{skillman98,kim09,kato13}, it is unclear whether this simplified picture holds. 

Negative superhumps, observed as (quasi-)coherent modulations with frequencies a few percent higher than the binary orbital frequency, reveal the presence of structures in the binary that move in a retrograde manner with respect to the orbital motion. The photometric modulations are the result of asymmetries in the binary plane which in the case of negative superhumps arise from the disk plane being offset from the orbital plane. Negative superhumps can be understood as beats between a retrogradely precessing disk and the prograde motion of the binary \citep{patterson99_negSH}, such that $f_{\rm negSH} = f_{\rm orb} - f_{\rm nodal}$, where $f_{\rm negSH}$ and $f_{\rm orb}$ are the negative superhump frequency and orbital frequency respectively, while $f_{\rm nodal}$ is the backward precession of the line of nodes of a misaligned and tilted disk. 

Initially thought to be the exception, the observational case for almost ubiquitous negative superhump signals in NLs has grown substantially. The main driver for this are the exquisite precision photometric lightcurves obtained by the \textit{Kepler} and \textit{TESS} satellites on AWDs. In several systems both the nodal precession frequency and its beat with the orbit (the negative superhump) are observed. In a handful of systems, negative superhumps even coexist with positive superhumps (the apsidal, prograde precession signal of an eccentric disk). Several harmonics associated with $f_{\rm negSH}$ and $f_{\rm nodal}$ are also regularly observed \citep[e.g.][]{ilkiewicz21,bruch23}.

Several authors were able to successfully reproduce period deficits and light-curve morphologies related to negative superhumps through the use of smooth-particle hydrodynamic (SPH) simulations \citep{wood00,montgomery09,wood09,montgomery12a,montgomery12b}, to the point where the disk tilt angle can be inferred from measurements of $f_{\rm negSH}$ and $f_{\rm orb}$. For a thin disk subject to the companion’s tidal field, a small global tilt can precess almost as a rigid body in the retrograde sense, producing the observed beat signal in SPH simulations. Analytical and numerical studies of tidally forced precession show that the precession period and the implied superhump period deficit $\epsilon_{\rm -} = (P_{\rm negSH} - P_{\rm orb})/P_{\rm orb} < 0$ scale with mass ratio and disk size in the way inferred from data compilations. Indeed, SPH studies provide empirical $\epsilon_{\rm -} (q)$ relations that are widely used to interpret observations \citep{montgomery09}.

However, although well modelled and phenomenologically understood, the big question remains: what causes the disk to tilt and what sustains this tilt in CVs in the first place? Clearly, a torque is required to lift the disk off the orbital plane, but also the disk itself needs to behave nearly as a rigid body (at least across the radii that dominate the optical modulation) to ensure that differential precession does not quickly smear out the signal. In practice, this means two conditions must be met: (i) there must be a steady source of vertical (out-of-plane) torque to excite and maintain the inclination, and (ii) there must be sufficiently rapid internal communication of that inclination, through viscosity and/or bending waves, such that the outer disk precesses coherently rather than twisting apart.

One possibility may be that the gas stream regularly overflows the rim or impacts obliquely, where its vertical momentum can pump the tilt at least once per orbit, with the hotspot acting as a moving force that is naturally phased to the observed beat. In high mass transfer rate NL variables, where the rim is hot and puffed up, overflow and a migrating impact point may be expected, making a stream-driven mechanism attractive. Once a small inclination exists it could naturally produce retrograde nodal precession at the observed frequencies and help regulate the tilt amplitude. Somewhat related to this, the disk need not to necessarily be symmetric between the ``top'' side relative to the ``bottom''. Through this asymmetry \cite{montgomery10} have proposed how a \textit{lift}, resulting from differing gas stream supersonic speeds over and under an accretion disk, could provide enough torque as it rotates to tilt the accretion disk and maintain it tilted.

A second route could involve weak magnetic stresses from a WD dipole field to the inner disk, which if misaligned with respect to the disk plane can exert a periodic vertical torque on the inner annuli. This inner torque could be communicated outwards in the disk providing a warp \citep[possibly through a wind as suggested for ultraluminous X-ray sources; see][]{middleton19}. However we expect this mechanism to require that the outward—provided warp propagates across the disk faster than nodal precession period or else the global disk tilt may not be achieved to reproduce the observed nodal precession. Furthermore, this route would require the WD magnetosphere to spin retrogradely relative to the disk in order to communicate retrograde precession, and it is not clear if this can be sustained.

A third route may involve the presence of a distant third body in a hierarchical configuration, whose inclined orbit exerts a secular gravitational torque on the accretion disk \citep[somewhat akin to the Kozai–Lidov cycles:][]{kozai62,lidov62}. Even a relatively low-mass companion, if sufficiently close and misaligned with respect to the inner binary plane, can induce a global tilt of the disk and drive retrograde nodal precession through classical three-body dynamics. Although the occurrence of triple systems is not negligible \citep[e.g.][]{toonen20,court19,martinez00,north00,knigge22,chavez22,shariat25}, it is not clear that \textit{all} AWDs displaying negative superhumps are hosting distant orbiting third bodies. After all, the prevalence of negative superhumps is substantially higher in NLs compared to DN system. This might argue against the third body explanation to induce retrograde nodal precession as NLs are thought as an evolutionary phase in CVs.

We'd also like to comment on a fourth route where disk winds (either magnetically, radiatively, or thermally launched) may be emitted asymmetrically from the disk. If the outflow geometry is even slightly tilted with respect to the orbital axis (either due to local instabilities, anisotropic irradiation, or coupling with the secondary’s wind) the resulting back-reaction torque could be enough to lift and maintain the disk inclination. The disk tilt may also be self-regulating such that a feedback loop between the wind launching region and the changing disk orientation might be at play. This mechanism could be particularly appealing in systems with strong outflows (see Section \ref{AWD_winds}), where even a modest asymmetry in the mass-loss rate could provide the necessary lever arm to sustain the observed long-term retrograde precession. 

Finally, we note that while instability to radiative warping \citep[e.g.][]{pringle96} is a known mechanism for tilting disks in high-luminosity systems (like AGN and X-ray binaries), it is unlikely to be effective in AWDs \citep[e.g.][]{murray98,foulkes10}. In these systems, the instability growth timescale is generally expected to be longer than the viscous timescale, preventing the warp from growing significantly.

Overall, a coherent model explaining the large prevalence of retrograde nodal disk precession in NLs is still required, and we strongly encourage others to tackle this challenge. We already know that this precessing tilted disks are common in AWD, and it is natural to expect the same to be true in other stellar-mass compact accretors. In fact, some low-mass X-ray binaries are known to undergo retrograde nodal disk precession \citep[e.g.][]{retter02,cornelisse13}. Ultimately, determining which (if any) of the  mechanisms outlined above dominate will require progress on both observational and theoretical fronts. Improved observational constraints will probably require time-resolved observations of disk wind, hot spot and/or magnetic field signatures. On the theoretical side, simulations designed to capture the physics of the most promising mechanisms -- e.g. the mass transfer stream from the donor star and/or the back reaction of an outflow on the disk -- are more likely to provide significant new insights.


\section{Do AWDs contain a hot ``corona''?}\label{sec:hotFlow}

We think a central question for accretion in AWDs is whether these systems host a geometrically thick ``hot flow'' (also referred to as a ``corona'') during episodes of quiescence, analogous to that inferred in X-ray binaries. It is unclear whether this additional ``hot flow'' component may, or may not, be present in addition to the classical SS73 thin disk and boundary layer. It is also unclear if, at least during quiescence, the boundary layer and the hot flow may be one and the same component, noting that for CVs in quiescence the classical optically thick boundary layer may in fact not be present at all \citep{dubus24}. Theoretically the hot flow was first introduced by \cite{narayan93} who found that the boundary layer transitions from optically thick to optically thin during low mass transfer rate episodes.

After all, all compact interacting binaries undergoing disk-driven outbursts are explained through the same thermal-viscous instability model, and XRBs undergoing outbursts display clear ``state changes'' where the inner region of the classic Shakura-Sunyaev accretion disk is replaced by an advection dominated accretion flow (ADAF) which is both geometrically thick and optically thin \citep{done07}. From a theoretical point of view, ``evaporation'' of the inner disk in AWDs has been postulated \citep{meyer94}, and truncation of the geometrically thin inner-disk is required to drive DN outbursts through the disk instability model \citep{dubus24}. So, does the transition where the inner disk becomes truncated and is replaced by a hot flow during low mass accretion rate episodes also happen in AWDs? If not, why not? In XRBs both the truncated classic Shakura-Sunyaev disk and the hot flow can be observed in X-ray spectra $\geq 10$\,keV during quiescent intervals \citep[the ``low/hard state''][]{done07}. Conversely from the blackbody geometrically thin disk radiation, the hot flow in X-ray spectra is observed as an inverse-Compton component, where electrons up-scatter disk photons to higher up to $\sim$100\,keV energies. During the DIM outbursts this hot flow is suppressed, and the classic geometrically thin disk pushes its way down the gravitational potential to reach the accretor \citep{lasota08}.

Given the relative large size of WDs compared to NSs and stellar-mass BHs, the gravitational potential energy release in the close proximity of AWDs is not large enough to heat the inner disk regions to X-ray temperatures (especially during quiescence). Consequently we would not expect to observe any comptonised component at X-ray wavelengths if a hypothetical hot flow were present to up-scatter disk photons. Although  the presence of a X-ray comptonised corona has been recently claimed \citep{maiolino20,titarchuk23}, it might be expected at extreme ultraviolet (EUV) wavelengths. Unfortunately we are blind to this window of the electromagnetic spectrum due to Galactic extinction and a lack of observing facilities, prohibiting us to directly test whether hot flows akin to those observed in XRBs exist in AWDs for the time being \citep{poutanen96}. 

Additionally to the direct evidence from the hard X-ray inverse-Compton emission, XRBs also manifest somewhat indirect evidence for the existence of hot flows. Over 2 decades ago \textit{RXTE} provided unprecedented X-ray lightcurves revealing large amplitude ($\approx 30\%$) aperiodic variability from minute to sub-second timescales. Further analysis has also revealed obvious features in the power spectral density (PSD) such as characteristic frequencies ($0.1 - 10$Hz) above which the variability power is suppressed. Furthermore it was clear early on that the X-ray timeseries shared a ubiquitous statistical phenomena such as the linear rms-flux relation and log-normal flux distributions. The ensemble of these timing studies results have been well explained by the so-called propagating fluctuations accretion disk model \citep{lyubarskii97,arevalo06,ingram11,ingram12,ingram13}, where disk material propagates inwards through the disk while being modulated on the \textit{local viscous timescale} which becomes faster as material moves deeper in the gravitational potential well. Beyond an inner-disk edge (be it geometrically thick or thin) there is no more disk to generate the mass transfer variations and as such the rms variability amplitude beyond a characteristic frequency (being the viscous frequency at the inner edge) will be heavily suppressed. The \textit{amplitude} of variations and the level of dilution of this variability as it propagate inwards is expected to be heavily dependent on the type of accretion flow. Geometrically thin disks can only provide small amplitude fluctuations on the order of a few percent, and these fluctuations are easily damped as they travel inwards. On the contrary, geometrically thick (and optically thin) flows can generate large amplitude variations which are not easily damped as they travel inwards. Because of this, and because of the inferred parameters from fitting X-ray power spectra, the large amplitude variations associated with the characteristic break frequency have generally been associated to the inner hot flow in XRBs \citep{done07}.

All of the timing phenomenology observed in XRBs have found a direct analogue at optical wavelengths in AWDs. Possibly the first AWD where the connection to XRBs was made is MV Lyrae, a novalike system that is practically face-on with respect to us, which was observed continuously by \textit{Kepler} for four years at 1 minute cadence. This relatively bright high mass transfer system, coupled with the long monitoring and photometric precision provided by \textit{Kepler} allowed to determine that its optical variability follows very similar phenomenology as XRBs \citep{scaringi12a}. The power spectral density (PSD) of this system shows a characteristic break at $\approx 10^{-3}$Hz \citep{scaringi12b}, a characteristic feature later also identified in several other AWDs \citep[e.g.][]{vandesande15}. Applying the same fluctuating accretion disk model to MV Lyrae provided constraints on the geometry of the variable emitting component, yielding $\alpha({H \over R})^2$ of order unity with a radial extent up to a few WD radii. Taken at face value, this would imply a geometrically thick inner disk \citep{scaringi14}. The challenge with MV Lyrae is that it is a high mass-transfer NL system, where one might naively expect the geometrically thin disk to push all the way to the WD surface, potentially quenching the variability and producing an optically thick boundary layer at the intersection of the WD surface and geometrically thin accretion disk. Yet this is not what is observed. 

This raises several questions, all related to the existence, or not, of this hot flow:
\begin{itemize}
    \item Are we seeing the boundary layer ``flicker'' instead of it being a hot flow? If so, how can such a small region produce such high amplitude variability and on such a wide range of timescales? Or is the boundary layer and the hot flow the same component with transitioning behaviour between accretion states?
    \item If a geometrically thick component does exist, then it must be optically thin, or else it would emit as a blackbody and we would have observed it by now. In this case, what process energises the electrons and sustains them?
    \item If the hot flow is optically thin and geometrically thick, why are we seeing large amplitude variability at optical wavelengths?
\end{itemize}

\noindent These are all big ``ifs'' that require direct observational evidence to be accepted or refuted. 

Some authors have instead interpreted the same PSD breaks in AWDs as the \textit{dynamical} (Keplerian) frequency of the inner edge of a geometrically thin disk \citep{revnivtsev11,balman12,shaw20} by inferring the truncation radius from:
\begin{equation}
    f_{\rm dyn}(R) = \frac{\Omega_{\rm dyn}(R)}{2\pi} = \frac{1}{2\pi}\sqrt{\frac{GM}{R^3}}
\end{equation}
This approach has further been applied to IPs showing breaks, where the inferred truncation radius is adopted to calculate the gravitational potential energy release in the polar caps of the magnetised WD, yielding the WD mass \citep{revnivtsev11,balman12,shaw20}. 
This is a tempting interpretation, but it is important to understand that it is inconsistent with the ``standard'' fluctuating propagation model. It is a fundamental assumption of this model that the variability is modulated on the \textit{viscous} timescale, approximated by:
\begin{equation}
    f_{\rm visc} = f_{\rm dyn}\alpha({H \over R})^2
\end{equation}
From this it is trivial to see that $f_{\rm dyn}\approx f_{\rm visc}$ for $\alpha ({H \over R})^2$ close to unity (expected for geometrically thick hot flows), making it difficult to interpret the origin of the frequency breaks from the PSD shape alone. If the breaks are the result of dynamical interactions, the open question is what generates the radiation? Orbiting material by itself does not radiate, bringing us back to the viscous interpretation of the breaks.

Establishing the existence or not of these hot flows in AWDs may have important implications for understanding accretion flows in other types of systems as well. The possibility that both XRBs and AWDs (and by extension AGN) share the same scale-invariant accretion disk physics is incredibly appealing (see Section \ref{sec:scaling}), but this connection is still far from being established. If hot flows of the sort seen ubiquitously in XRBs are really present in AWDs, such a link would 

With this in mind, we hope to have laid out the observational evidence for hot flows in AWDs, and the fundamental challenges in testing the existence of these flows in these systems.

\section{Where and how are disks truncated by the effects of spinning magnetospheres in AWDs?}

Magnetic fields of accreting objects are of key importance in the accretion process influencing both geometry and emission processes. While strong magnetic field WDs ($>$ 10-20\,MG) observed in polar-type CVs prevent the formation of an accretion disk, here we deal with systems where the $B$-field strength is not high enough to entirely inhibit disk formation but rather truncate its inner regions, namely IPs (and possibly some NLs). 

The dynamics of the accretion flow as it approaches the WD is expected to be determined by the magnetic interaction occurring at the magnetospheric boundary where the magnetic pressure balances the ram pressure of the infalling matter \citep{FKR}. The IP-type CVs are believed to harbour moderately magnetic WDs ($\lesssim$ 10\,MG) as they display fast (tens of sec - min) variations from X-rays to optical bands ascribed to the rotation of the accreting magnetized WD. Their asynchronism with respect to the orbital period (hours) makes them distinct from the polars whose WD rotation is locked at the orbital period \citep[see][]{cropper90,warner03}. It is relatively well established that the accretion flow from the disk magnetospheric boundary is channelled onto the WD magnetic poles in an arc-shaped curtain \citep{rosen88}. However where the disk is truncated is not easy to assess. 

When thinking about how disks around WDs behave, three characteristic radii often come up in conversation: the circularisation radius $R_{\rm circ}$, the magnetospheric (Alfv\'en) radius $R_{\rm m}$, and the corotation radius $R_{\rm co}$. Which one dominates tells us a lot about whether a disk forms at all, how far it extends, and whether material actually makes it onto the WD. The circularisation radius is approximately defined as,
\begin{equation}
\frac{R_{\rm circ}}{a} \;\simeq\; (1+q)\,\left[\,0.5 - 0.227\,\log q\,\right]^4,
\end{equation}
where $a$ is the binary semi-major axis and $q={M_2 \over M_1}$ the mass ratio, and simply defines the radius at which the incoming stream from $L_1$ would settle if it conserved angular momentum. As long as $R_{\rm circ}$ is outside the WD’s surface, a disk can in principle form. For typical CV mass ratios this is comfortably the case. Magnetic fields complicate matters. The magnetospheric (Alfv\'en) radius sets the point where magnetic stresses balance the ram pressure of the accretion flow,
\begin{equation}
R_{\rm m} \;=\; \left( \frac{\mu^4}{2 G M \dot{M}^2} \right)^{1/7},
\end{equation}
with $\mu = B R^3$ the dipole moment. If $R_{\rm m}$ lies outside $R_{\rm circ}$, the magnetic field intercepts the stream before a disk can even form: accretion is then “stream-fed,” as in polars. If $R_{\rm m} < R_{\rm circ}$, a disk can exist, but it is truncated on the inside by the magnetosphere, as in the IPs. Finally, the corotation radius tells us where the disk matches the WD spin ($P_{\rm spin}$),
\begin{equation}
R_{\rm co} \;=\; \left(\frac{G M P_{\rm spin}^2}{4\pi^2}\right)^{1/3}.
\end{equation}
If $R_{\rm m} < R_{\rm co}$, the magnetosphere rotates more slowly than the Keplerian disk and material can funnel in along the field lines. If $R_{\rm m} > R_{\rm co}$, the magnetosphere whirls faster than the disk -- the dreaded \emph{propeller regime} -- and matter is more likely to be expelled than accreted. So the fate of the flow in magnetic CVs boils down to the ordering of these radii:
\begin{itemize}
  \item $R_{\rm circ} < R_{\rm m}$: no disk, stream-fed accretion (e.g. polars).
  \item $R_{\rm m} < R_{\rm circ}$ and $R_{\rm m} < R_{\rm co}$: truncated disk accretion (most IPs).
  \item $R_{\rm m} > R_{\rm co}$: propeller effect, with partial or no accretion (e.g. the IP AE Aqr).
\end{itemize}

This simple set of inequalities makes clear why AWDs are such handy laboratories: by measuring or constraining $P_{\rm spin}$, $B$, and $\dot M$, we can literally ``dial in'' which flavour of accretion a given system should exhibit. In practice, nature seems to populate all these regimes giving us a zoo of CV behaviours that can attempt to map back to this trio of radii. If $R_{\rm mag}<R_{\rm co}$ then disk material latching onto the WD magnetic field lines will transfer angular momentum to the WD and spin it up in the process. Conversely if $R_{\rm mag}>R_{\rm co}$ disk material will instead be propelled away, carrying with it angular momentum extracted from the WD spin (as the $B$ field is assumed to be anchored to the WD surface), resulting in the WD spinning down. In general it is thought that the WD in IPs tend to find an equilibrium where $R_{\rm mag}\approx R_{\rm co}$ \citep{norton04,norton08}.

As introduced in Section \ref{sec:hotFlow} the presence of a frequency break in the X-ray and optical bands, although difficult to measure, has been recently used to derive the radius at which the disk is truncated in IPs \citep{suleimanov19}. This assumes that the maximum possible magnetospheric radius is equal to the corotation radius and that the frequency break traces the dynamical Keplerian frequency at a specific disk radius. As discussed in Section \ref{sec:hotFlow}, the observed frequency breaks are most likely related to the \textit{viscous} frequency rather than the dynamical Keplerian one, somewhat questioning this methodology. Furthermore the assumption that the magnetospheric and co-rotation radii are the same is not always a valid one since the radius at which truncation occurs depends not only on the magnetic field strength but also on the accretion rate \citep{revinivtsev09}. Changes in the accretion rate can also affect the location of the magnetospheric radius, moving it inward or outwards and thus IPs observed at different luminosity levels may help in deriving changes in the magnetospheric radius. This in turn makes WD mass determinations by using the magnetospheric radius derived from break frequencies rather uncertain. An additional important aspect is that the accretion rate also affects the shape of the post-shock region \citep[PSR;][]{aizu73} which at low rates can be tall and diverge, also affecting the WD mass determination \citep{suleimanov25}.


The dichotomy ``magnetic'' vs. ``non-magnetic'' has historically been an observational one alone, but it is natural to expect that WDs possess a continuum of magnetic field strengths, even below the canonical $\approx 10^6$G adopted for a system to be ``magnetic''. Some systems, conventionally classified as ``non-magnetic'' such as some NLs, have shown observational phenomenology of subtle effects that have been associated to the WD magnetic field interaction with the disk. 

The NL MV\,Lyrae when caught in a low state by \emph{Kepler} exhibited bursts recurring quasi-periodically every 2\,h lasting about 30\,min and interpreted as magnetically gated accretion episodes from the accumulation of matter at the magnetospheric radius of a weakly ($<10^6$ G) magnetic WD \citep{spruit93,dangelo12,scaringi17,scaringi22c}. Similar phenomenology was observed in another NL system, TW\,Pictoris, with \emph{TESS}, additionally exhibiting even more complex variability patterns, transitioning rapidly from high to low states on timescale as short a 30\,min and switching back to high states on a longer timescale of $\sim$12\,h. This behaviour is phenomenologically very reminiscent of so-called mode switching in transitional millisecond pulsars binaries, albeit on longer timescales \citep[as may be expected from the larger WD accretor compared to a NS, see e.g.][]{scaringi22c}. While in transitional millisecond pulsars the mode switching has been  interpreted as discrete mass ejection episodes from multi-wavelength campaigns \citep{baglio23}, the mode transitions in TW Pic still await full multi-band observations to test this hypothesis.

Other potential explanations have also been put forward, in particular one where changes in the accretion rate or a reconfiguration of the magnetic field geometry drives a propeller \citep[rather than accretion,][]{scaringi22c}. These possibilities however raise the question as to what causes the accretion rate and/or magnetic field reconfiguration to be altered. Interestingly during the ``low-mode states'' TW\,Pictoris displayed bursts lasting about 30\,min, which very much resemble those observed in MV\,Lyrae, suggesting that magnetic gating preferentially occurs at low-levels of accretion. From these early studies the question of how many NLs display this type of variability (and in turn how do magnetic field strengths and mass accretion rates determine this behaviour) remains an open one. The similarity of magnetically gated bursts as well as the mode switching observed in weakly magnetised NS binaries, although on different timescales, opens a new window in studying magnetic field configurations over a large range. Given both these systems have been identified through space-based photometric monitoring missions (\textit{Kepler} and \textit{TESS}), it is natural to expect that more may be discovered through similar methods, and in this respect the upcoming \textit{PLATO} \citep{plato} mission may be fundamental in understanding this poorly studied accretion mode. 

Finally, we'd like to comment on a more recent result from \cite{veresvarska24b} which have found a handful of AWDs displaying persistent quasi-periodic oscillations (QPOs) on timescales of $\approx80$ minutes through long term monitoring using \textit{TESS} data. Interestingly, these QPOs have been directly compared to so called Type-C QPOs observed in accreting stellar-mass black holes and horizontal brand oscillations (HBOs) observed in accreting neutron stars \citep[see][for a review of QPOs in XRBs]{ingram19}. QPOs of this type in XRBs have generally been associated to general relativistic frame-dragging of an inner hot-flow causing the inner regions to undergo so-called Lense-Thirring precession \citep[][]{stella98}. Given AWDs are inherently classical systems, general relativistic effects are not large enough to affect the accretion flow dynamics. However, a hypothesis to explain these QPOs in AWDs has been put forward, where a weak WD dipole magnetic field ($B\approx10^4 - 10^5$G) is misaligned to the disk plane and is able to sustain a magnetically-driven inner disk warp \citep[][]{lai99,pfeiffer04}. Although this interpretation hinges on several model assumptions, it is interesting that it can still explain the observed QPO phenomenology in AWDs and possibly NS-XRBs and pulsating ultraluminous X-ray sources \citep{veresvarska25b}. For AWDs we note that, although the sample is currently limited to 5 systems, 2 of these are some of the closest AWDs (WZ Sge and GW Lib), suggesting that QPOs in AWDs may be more common than previously thought. Testing the magnetically-driven precession hypothesis does not appear trivial due to the large uncertainties caused by the several model assumptions. One way forward may be to understand if MHD simulations of weakly magnetised AWDs can drive and maintain precessing inner flows when a weak dipole is misaligned to the disk plane.

\section{What triggers outbursts in Intermediate Polars?}

Outbursts in AWDs, and specifically in CVs, are common, with up to 40$\%$ of systems undergoing DN outbursts. This is very different in IPs, which rarely show outbursts \citep{hellier99,hameury17}. As the accretion disks in IPs are generally truncated at the inner-edges by a rotating magnetosphere, there have been several studies debating whether it is possible for the DIM to trigger DN outbursts in the few IPs that do show such events since these outbursts are relatively short compared to classic DNe. Alternative explanations involve mass transfer enhancements from the donor \citep{hameury17}. 

\begin{figure*}
\includegraphics[width=1\textwidth]{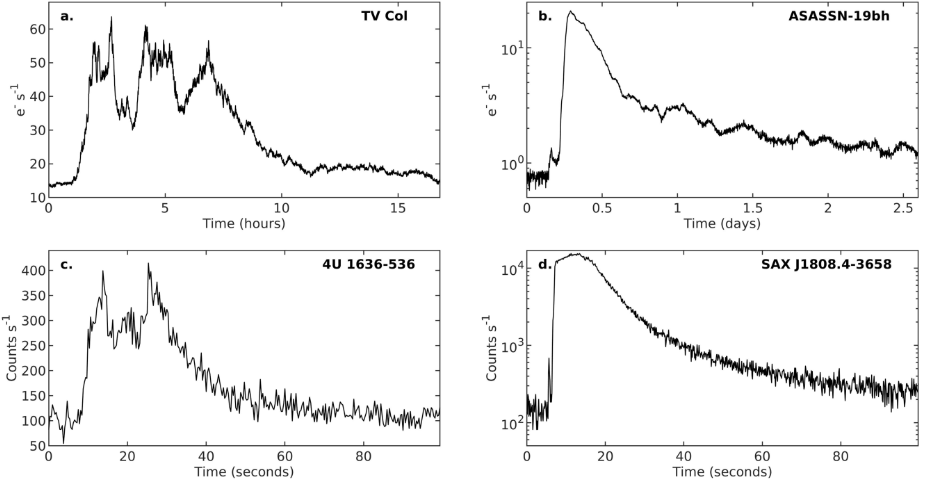}
\caption{ Comparison between Type-I X-ray bursts and micronovae. The top two panels (a \& b) show \textit{TESS} lightcurves the rapid bursts observed in TV Col and ASASSN$-$19bh \citep{scaringi22a}. The bottom two panels (c \& d) show X-ray lightcurve of 4U 1636$-$536 observed with \textit{EXOSAT}-ME and SAX J1808.4$-$3658 observed with \textit{RXTE}-PCA X-ray. Similarities in the lightcurve morphology between the TV Col and 4U 1636-536 (e.g. multi-peaked burst) and between ASASSN$-$19bh and SAX J1808.4$-$3658 (e.g. pre-cursor) are obvious. \citep[Figure credit:][]{scaringi22a}}
\label{fig:micronova}
\end{figure*}

More recently observations of IPs with the \emph{TESS} satellite have revealed an increasing number of IPs displaying energetic bursts up to a few magnitudes that rise relatively fast (tens of minutes) and exponentially fade over less than a day \citep{scaringi22a,scaringi22b,ilkiewicz24,irving24,veresvarska25a}. The shape of these bursts finds a remarkable similarity to so-called Type-I X-ray bursts routinely observed in some accreting NSs (see Fig. \ref{fig:micronova}) and explained as thermonuclear runaway explosions of freshly accreted material onto the NS surface \citep{galloway08}. The energetics and timescales suggest that these events could be due to localised thermonuclear runaway (TNR) explosions from the small surface of the polar region of the magnetic WD where accretion takes place. These events have been dubbed ``micronovae'' since the energy release is $\sim 10^{-6}$ times lower than those of classical novae \citep{scaringi22b}. However it is entirely unclear how, if, and for how long magnetic confinement can sustain material within the accretion column for prolonged enough time. Different interpretations have also been put forward for the short outbursts in IPs: magnetic gating may be an option for systems like V1223\,Sgr \citep{hameury22}, and it may be that both positive superhumps and short outbursts are somewhat linked as in the case of SU UMa-type DNe \citep{mukai23}. It has also been noted that the classification of outbursts is not straight forward \citep[see][]{ilkiewicz24}, but diagnostics such as burst duration, luminosity and energy can be a useful proxy in distinguishing between micronovae, magnetic gating and DNe. Increasing the observed samples of outbursts in IPs and potentially simultaneously monitoring these bursts at X-ray wavelengths will allow to test the TNR hypothesis for micronovae. We conclude by stating that a TNR origin for micronovae (or a magnetic reconfiguration origin) are hypothesis that require not only observational tests, but importantly simulations of such events that can provide physical intuition and testable predictions to rule-out or accept the TNR scenario for these events.

\section{Can AWDs be used as universal accretion laboratories?}\label{sec:scaling}

If for a moment we only focus on phenomenology and attempt to not seek a physical validation for our observations we will find that AWDs behave in a remarkably similar way to XRBs, and by extension possibly to active galactic nuclei (AGN) and young stellar-objects (YSOs). Specifically the timing properties and the observed high frequency breaks (see Section \ref{sec:hotFlow}) in accreting objects appear to all be self-similar once scaled to the size of the accretion disk (or equivalently to the size of the accreting object in question).

The fundamental observation which appears to best support disk scale-invariance across scales in a quantitative manner is the so-called accretion-induced variability plane. Originally discussed in the context of XRBs and AGN \citep{mchardy06}, this plane links the observed PSD break frequencies to the mass of the accretor and the mass transfer rate, such that:

\begin{equation}
f_{\rm b} \propto M^{-2} \dot{M}.
\label{eq:scale1}
\end{equation}

\noindent The relation in Eq. \ref{eq:scale1} has been shown to hold for accreting BHs (both stellar-mass and supermassive) across 8 orders of magnitude in mass. The variability plane has been extended to include AWDs and possibly YSOs \citep{scaringi15}. The inclusion of AWDs enabled to break a degeneracy which has plagued BH-only samples. This is because the inner-most disk radius for BHs in the high/soft state, where the geometrically thin disk is thought to push all the way down to the accretor surface/event horizon, linearly scales with the mass of the accretor (and BH spin to a lesser degree). Including AWDs to the sample somewhat breaks this degeneracy, to find that:

\begin{equation}
f_{\rm b} \propto R_{\rm in}^{-2} \dot{M},
\label{eq:scale2}
\end{equation}

\noindent where $R_{\rm in}$ is the inner-most disk radius. 

Although the accretion-induced variability plane has been extended to include AWDs and possibly YSOs, the physical reasons as to why this scaling appears to hold within specific systems, or across the scales, remains elusive. Furthermore, although the break frequency $f_{\rm b}$ has been shown to linearly depend on $\dot{M}$, the exact dependence on $R_{\rm in}$ and $M$ has not been so strongly established. Although including AWD to the sample of accreting BHs yields a relation as that in Eq. \ref{eq:scale2}, a dependence on $M$ is strictly not ruled out. The strongest constraints from both the \cite{mchardy06} and \cite{scaringi15} results on both $R_{\rm in}$ and $M$ are that the sum of the exponents on both $R_{\rm in}$ and $M$ is $-2$. Thus in general

\begin{equation}
f_{\rm b} \propto R_{\rm in}^\alpha M^\beta \dot{M}
\label{eq:scale3}
\end{equation}  

\noindent with $\alpha + \beta = -2$. The empirical scaling relation described here offers no true physical interpretation, but it is remarkable that all these very different accreting systems appear to follow it. Is this a ``cosmic coincidence'' or is there real physics to be learned here?

In the spirit of (re-)starting a debate on this we'd like to be provocative, and put forward a phenomenological prescription, purely based on phenomenology, that strikingly appears to hold across all accreting systems. Our intuition is grounded on the constrain inferred from Eq. \ref{eq:scale3}, which we use in combination with dimensional analysis to put forward the provocative ansatz that the ratio of break frequency to dynamical frequency ($f_{\rm b} \over f_{\rm dyn}$) is directly proportional to the Eddington ratio $L_{\rm acc} \over L_{\rm edd}$:

\begin{equation}
    {f_{\rm b} \over f_{\rm dyn}} = {L_{\rm acc} \over L_{\rm edd}} = {\dot{M} \over \dot{M}_{\rm edd}} .
	\label{eq:1}
\end{equation}

\noindent where $f_{\rm dyn}$ is the dynamical Keplerian frequency at radius $R_{\rm in}$ and $L_{\rm acc}={1 \over 2} {GM\dot{M} \over R}$ is the accretion luminosity. This type of scaling may be expected for relatively high Eddington ratios, but it is entirely unclear why this may hold at very low ratios. Nonetheless, under this assumption, the break frequency $f_{\rm b}$ can be expressed as:

\begin{equation}
f_{\rm b} = {{\epsilon c \sigma_{\rm T}} \over {8 \pi^2 m_{\rm p} \sqrt{G}}} M^{-1 \over 2} R^{-3 \over 2} \dot{M},
\label{eq:2}
\end{equation}

\noindent where $\sigma_{\rm T}$, $c$, $m_{\rm p}$, $G$ are the Thompson cross section, speed of light, proton mass, and Gravitational constant respectively, and $\epsilon$ is the radiation efficiency for accreted material to be converted into radiation at the Eddington limit via $L_{\rm edd} = \epsilon c^2 \dot{M}_{\rm edd}$. 

In general it can be assumed (to zeroth order) that the conversion efficiency $\epsilon$ is directly related to the conversion of gravitational potential energy into radiation such that $\epsilon = {{GM} \over {Rc^2}}$. For accreting BHs $\epsilon \approx 0.1$ whilst for accreting WDs $\epsilon \approx 10^{-3}$. In general however, we can rewrite Eq. \ref{eq:2} using the definition of $\epsilon$,

\begin{equation}
f_{\rm b} = {{\sigma_{\rm T} \sqrt{G}} \over {8 \pi^2 m_{\rm p} c }} M^{1 \over 2} R^{-5 \over 2} \dot{M}.
\label{eq:3}
\end{equation}

The dependence of $f_{\rm b} \propto  M^{-1 \over 2} R^{-3 \over 2} \dot{M}$ in Eq. \ref{eq:2} as well as the dependence of $f_{\rm b} \propto  M^{1 \over 2} R^{-5 \over 2} \dot{M}$ in Eq. \ref{eq:3} are both consistent with previously reported observed scaling relations of \cite{mchardy06} and \cite{scaringi15} such that the sum of the mass and radius exponents is $-2$. 

\begin{figure*}
\includegraphics[width=1\textwidth]{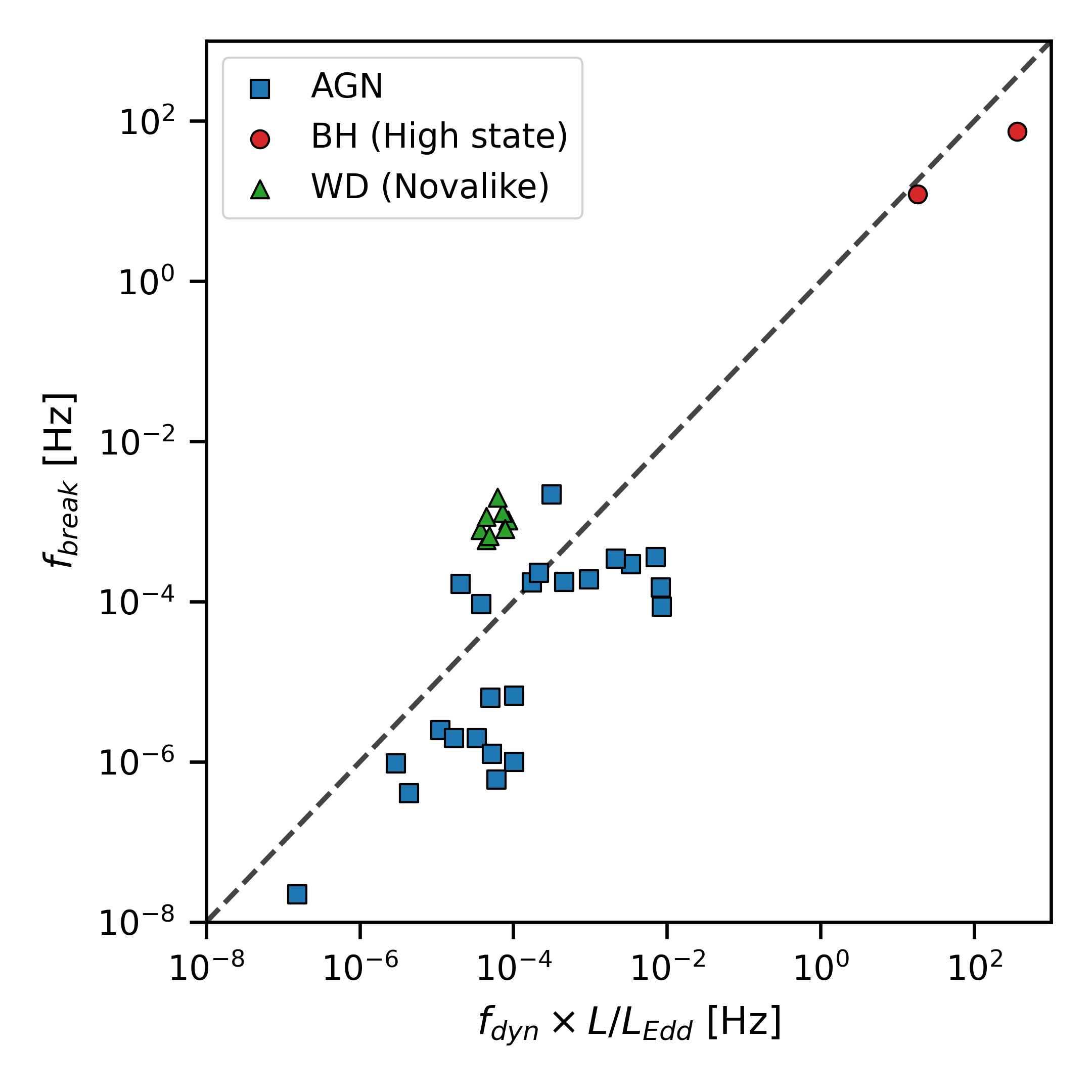}
\caption{Observed PSD break frequencies for accreting black holes and white dwarfs compared to the semi-empirical scaling from Eq. \ref{eq:3} (black line). The consistency across ten orders of magnitude in mass may suggest a near-universal relation linking variability timescales and accretion power. Figure adapted from \cite{scaringi15}.}
\label{fig:plane}
\end{figure*}

The relation in Eq.~\ref{eq:3} is plotted in Fig.~\ref{fig:plane} against the previously published values of PSD break frequencies in both stellar mass and supermassive BHs and accreting WDs \citep{mchardy06,scaringi15}. The thick black diagonal line is not a fit: on the contrary this shows how the simple assumption made in Eq.~\ref{eq:1} that was recovered from dimensional analysis describes the observations incredibly well. The AWDs and AGN shown in Fig.~\ref{fig:plane} appear to depart from the relation by $\approx$1~dex, albeit the observations stellar-mass of BHs appear to be consistent. Nonetheless it is somewhat remarkable that not only the exponents of the phenomenological scaling relation are recovered, but interestingly the \textit{offset} to the plane given by the constant in Eq.~\ref{eq:3} is also matching observations with a scatter across the relation: is this also a ``cosmic coincidence'', or is real physics waiting to be unravelled?

If we attempt to interpret Fig.~\ref{fig:plane} a few systematic effects deserve emphasis. First, the adopted $L_{\rm Edd}$ is the classical Thomson value, and in discs the relevant opacity can deviate from pure electron scattering (composition, ionization), and inclination-dependent anisotropy can bias $L_{\rm bol}$ by factors of a few. Second, the definition of what constitutes a PSD break is method–dependent (bending power law vs. Lorentzian join, windowing and red–noise leakage corrections), and heterogeneous choices can shift points vertically by $\sim$0.3–0.5~dex. Third, for AWDs a substantial fraction of the disk luminosity can emerge in the unobservable EUV band, so $L_{\rm acc}$ can easily be underestimated. Correcting for missing EUV can potentially move the NL AWD points rightward toward the relation. Finally, bandpass can in principle depress the optical break and shift it to lower frequencies if it is tied to local fluctuations at the radius dominating a given wavelength under the SS73 disc prescription. Once again we are left to postulate that the optical variability traces signatures of an inner-disk variability clock (see Section \ref{sec:hotFlow}), in which case bandpass effects could introduce modest, system–dependent offsets rather than a wholesale shift for all systems. 

We point out that AGN can display both optical and X-ray PSD breaks, and although the X-ray breaks have conventionally been interpreted as the viscous frequency at the inner-disk edge, the optical breaks have instead been interpreted as the thermal frequency at the disk emitting radius contributing to the observed pass-band \citep{burke21}. This could also be at play in AWDs, highlighting our very different interpretations to the frequency breaks observed in these systems. So, is the break frequency observed across WDs (and across compact accretors in general) a dynamical \citep{revnivtsev11}, viscous \citep{scaringi14} or thermal one \citep{burke21}? It seems all available options are on the table, highlighting how far we currently are from unravelling this puzzle.

Stepping back, the semi-empirical relation of Eq.~\ref{eq:3} across compact accretors suggests that the break is a timescale–power coupling anchored to the inner dynamical clock, rather than a coincidence. This may not seem surprising as all timescales of relevance (dynamical, viscous and thermal) are dependent on the local dynamical one. What remains somewhat surprising is that the ansatz in Eq.~\ref{eq:1} can be read as stating that the offset in relation scales with the Eddington-normalized accretion power across all systems, a postulate we currently have no physically grounded intuition for but hope others may explore in more detail.

\section{Conclusion}\label{sec13}
AWDs remain our most accessible and revealing laboratories for studying disk accretion. Yet, as this ``unreview'' highlights, even these seemingly simple systems continue to challenge our understanding. The origin of disk viscosity, winds, disk tilts, truncation radii, and outbursts are all phenomena which still lack a depth of understanding required to fully explain. Addressing these open questions will not only clarify the physics of AWDs themselves but also refine our broader models of accretion across all mass scales. To finally conclude, we hope to have laid the ground work for future keen scientists to tackle these challenges and hopefully help unravel their physical origin and/or raise new hypotheses to test!

\backmatter

\bmhead{Acknowledgements}
SS acknowledges support by STFC grant ST/T000244/1 and ST/X001075/1.
CK acknowledge support by the Science and Technology Facilities Council grant ST/V001000/1.
DdM acknowledges support from the Italian Institute of Astrophysics INAF through ``Astrofisica Fondamentale 2022 - Large grant N.16''.
All authors acknowledge the International Space Science Institute (ISSI) for their support and organisation of the ``Accretion Disks: The First 50 Years'' workshop held in Bern.
Finally, all authors acknowledge the dedication and inspiration set out by Brian Warner, Tom Marsh, Tomaso Belloni, Jean-Pierre Lasota and Maurizio Falanga, all of whom have shaped the field of accretion physics and accreting white dwarfs for decades to come. This ``unreview'' would have not been possible without their leading intuition, infectious enthusiasm, and the massive foundations they laid for the field. We stand on their shoulders as we look for the answers to these open question.

\section*{Declarations}

The authors declare no competing interests.

\bibliography{sn-bibliography}

\end{document}